\documentclass[10 pt, aps,twocolumn]{revtex4-1}

\usepackage{graphics} % for pdf, bitmapped graphics files
\usepackage{epsfig} % for postscript graphics files
\usepackage{amsmath} % assumes amsmath package installed
\usepackage{amssymb}  % assumes amsmath package installed
\usepackage{amsfonts, color}
\usepackage{graphicx, subfigure}
\usepackage[normalem]{ulem}

\begin{document}

\title{Statistical multi-moment bifurcations in random delay coupled swarms }
\author{Luis Mier-y-Teran-Romero$^{(1)}$, Brandon Lindley$^{(2)}$,  and Ira B. Schwartz$^{(3)}$}
%\email{dykman@pa.msu.edu}
\affiliation{$^{(1)}$ US Naval Research Laboratory, Code 6792, Nonlinear System Dynamics Section, Plasma Physics Division, Washington, DC 20375}
\date{\today}

%%%%%%%%%%%%%%%%%%%%%%%%%%%%%%%%%%%%%%%%%%%%%%%%%%%%%%%%%%%%%%%%%%%%%%%%%%%%%%%%
\begin{abstract}
We study the effects of discrete, randomly distributed time delays on the
dynamics of a coupled system of self-propelling  particles. Bifurcation analysis on a
mean field approximation of the system reveals that the system possesses patterns with certain universal characteristics that 
depend on distinguished moments of the time delay
distribution. Specifically, we show both theoretically and numerically that although bifurcations of simple
patterns, such as translations, change stability only as a function of the
first moment of the time delay distribution, more complex patterns arising from
Hopf bifurcations depend on all of the moments. 

\end{abstract}

%%%%%%%%%%%%%%%%%%%%%%%%%%%%%%%%%%%%%%%%%%%%%%%%%%%%%%%%%%%%%%%%%%%%%%%%%%%%%%%%

\maketitle

% \section{Comments from IBS}
% The main results are the following as I see them:
% 1. We have changes of instability for simple bifurcations depending on
% the first moment only.
% 2. We more complicated patterns of rotational nature depending on all moment
% of the distributions. 
% The two items are our conclusions. I have drafted an abstract that better
% reflects the discoveries. 

Recently, much attention has been given to the study of interacting
multi-agent, particle or swarming systems in various natural and engineering
fields. Interestingly, these multi-agent swarms can self-organize and form complex spatio-temporal patterns even when the coupling between agents is weak. Many of these investigations have been motivated by a
multitude of biological systems such as schooling fish,
swarming locusts, flocking birds,  bacterial colonies, ant movement,
etc. \cite{Budrene95, Toner95, Parrish99, Topaz04}, and have also been applied
to the design of systems of autonomous, communicating robots or agents \cite{Leonard02, Morgan05, chuang2007} and mobile sensor networks \cite{lynch2008}.

Many studies describe the swarm system at the individual, or particle, level via models
constructed with ordinary differential equations (ODEs) or delay differential
equations (DDEs) to describe the  trajectories
\cite{vicsek95,flierl99}. When there are a large number of densely-distributed
particles, authors have employed partial differential equations (PDEs)
to describe the average agent density and velocity
\cite{Toner95,Toner98,Edelsteinkeshet98, Topaz04}. Recently, the inclusion of noise in such
particle-based studies has revealed interesting, noise-induced 
transitions between different coherent patterns \cite{Erdmann05,
  Forgoston08}. Such noise driven systems have led to the discovery of first and
second order phase transitions in swarm models~\cite{aldana07}.

A topic of intense ongoing research in interacting particle systems, and in
particular in the dynamics of swarms, is the effect of time delays. It is well
known that time delays can have profound dynamical consequences, such as
destabilization and synchronization \cite{Englert11, Zuo10}, and delays have
been effectively used for purposes of control \cite{Konishi10}.  Initially, such
studies focused on the case of one or a few discrete time delays. More
recently, however, the complex situation of several and random
time delays has been researched \cite{Ahlborn07, Wu09, Marti06}. An additional
important case is that of distributed time delays,  when the dynamics of
the system depends on a continuous interval in its past instead of on a discrete instant \cite{Omi08,Dykman12}.

There exists a complex interplay  between the attractive coupling, time
delay, and noise intensity that produces transitions between different
spatio-temporal patterns \cite{Forgoston08,MierTRO12} in the case of a
  single, discrete delay. Here, we
consider a more general swarming model where coupling information
  between particles occurs with randomly distributed time delays.  We perform
a bifurcation analysis of a mean field approximation and reveal the patterns
that are possible at different values of the coupling strength and parameters
of the time delay distribution.  

We  model the dynamics of a 2D  system of $N$ identical
self-propelling agents that are attracted to each other in a symmetric
manner. We consider the effects of finite communication speeds and 
information-processing times so that the attraction between agents occurs in a
time delayed fashion. The  time delays are non-uniform but they
are symmetric among agents $\tau_{ij}(=\tau_{ji})$, for particles $i$ and $j$,
as well as constant in time. The dynamics of
the particles is  described by the following governing equations:
\begin{align}\label{swarm_eq}
\ddot{\mathbf{r}}_i =& \left(1 - |\dot{\mathbf{r}}_i|^2\right)\dot{\mathbf{r}}_i -
\frac{a}{N}\mathop{\sum_{j=1}^N}_{i\neq j}(\mathbf{r}_i(t) -
\mathbf{r}_j(t-\tau_{ij})),
\end{align}
for $i =1,2\ldots,N$. The vector $\mathbf{r}_i$ denotes the
position of the $i$th agent at time $t$. The term
$\left(1 - |\dot{\mathbf{r}}_i|^2\right)\dot{\mathbf{r}}_i$ represents self-propulsion and frictional drag forces that act on each agent. The coupling
constant $a$ measures the strength of the attraction between agents and the time delay between particles $i$ and $j$ is  given by
$\tau_{ij}$. When $a=0$ the agents tend to  move in a straight line with unit speed as time tends to infinity. The $N(N-1)/2$ different time delays
$0<\tau_{ij}(=\tau_{ji})$ are drawn from a distribution $\rho(\tau)$ whose 
mean and standard deviation are denoted by $\mu_\tau$ and $\sigma_\tau$, respectively.

We obtain a mean field approximation of the swarming system by measuring the
particle's coordinates relative to the center of mass $\mathbf{r}_i = \mathbf{R} + \delta
\mathbf{r}_i$, for $i =1,2\ldots,N$, where  $\mathbf{R}(t) =
\frac{1}{N} \sum_{i=1}^N\mathbf{r}_i(t)$.  Following the approximations from
\cite{Lindley12}, we obtain a mean field description of the swarm:
\begin{align}\label{mean_field}                                                                  
\ddot{\mathbf{R}}=& \left(1 - |\dot{\mathbf{R}}|^2 \right)\dot{\mathbf{R}}
-a\left(\mathbf{\
R}(t) - \int_0^\infty \mathbf{R}(t-\tau)\rho(\tau)d\tau \right).                                                               
\end{align}
The approximations necessary to obtain Eq. \eqref{mean_field} require that
 $N$ be sufficiently large so that $\frac{1}{N(N-1)}\sum_{i=1}^N
\mathop{\sum_{j=1}^N}_{i\neq j}\mathbf{R}(t-\tau_{ij}) \approx
\int_0^\infty \mathbf{R}(t-\tau)\rho(\tau)d\tau$ and
that the swarm particles remain close together. 

\vspace{-0.3cm}

\begin{figure}[h]
\begin{minipage}{0.49\linewidth}
\includegraphics[width=4.3cm,height=3.0cm]{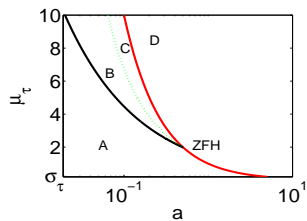}
\end{minipage}
\vspace{-0.3cm}
\caption{Bifurcation structure of the translating state of the mean field
  Eqs. \eqref{mean_field} for the exponential distribution described in the
  text ($\sigma_\tau=0.2$). The translating state merges with: (\emph{i}) the
  stationary state along the continuous red
  curve $a\mu_\tau=1$; and (\emph{ii})  the circularly rotating state along the
  dashed black curve $a \langle \tau^2 \rangle = 2$. The component of the translating state parallel to the
motion undergoes a Hopf bifurcation along the green, dotted curve. (Color online.)}\label{trans_bifs}
\end{figure}

\vspace{-0.2cm}

Equations \eqref{mean_field} admit a uniformly
translating solution $\mathbf{R}(t) = \mathbf{R}_0 + \mathbf{V}_0 \cdot
t$ ($\mathbf{R}_0$ and $\mathbf{V}_0$ are any constant 2D vectors). The speed $|\mathbf{V}_0|$ must satisfy
\begin{align}\label{speed_sqr}
  |\mathbf{V}_0|^2 = 
1-a\int_0^\infty \tau \rho(\tau)d\tau = 1 - a\mu_\tau,
\end{align}
 which shows that this solution is possible as long as the system parameters
lie below the hyperbola  $a \mu_\tau =
1$ in the $(a, \ \mu_\tau)$ plane.  Remarkably, the speed of the of the
translating state depends exclusively on the mean of the distribution
$\rho(\tau)$ and not on any of the higher moments.

 The linear
stability of the translating state is examined  by taking $X(t) = \sqrt{1 -
  a\mu_\tau} \cdot t + \delta
X(t)$ and $Y(t) = \delta Y(t)$. The two linearized equations decouple and the
stability of motions parallel and perpendicular to the translating direction
are determined by the characteristic equations
$\mathcal{D}_{\parallel}(\lambda)$ and $\mathcal{D}_{\perp}(\lambda)$, respectively:
\begin{align}\label{char_eq_par_perp}
\mathcal{D}_{\parallel}(\lambda) = \mathcal{F}(\lambda) - (3a\mu_\tau - 2)
\lambda, \quad \mathcal{D}_{\perp}(\lambda) = \mathcal{F}(\lambda) - a\mu_\tau \lambda,
\end{align}
where $\mathcal{F}(\lambda) = a\left(1- \langle e^{-\lambda\tau} \rangle
\right) + \lambda^2$. The function $\langle e^{-\lambda\tau}\rangle$ is the moment generating function of $\rho(\tau)$ since the
$n$-th moment is equal to $\langle \tau^n\rangle = (-1)^n\frac{d^n}{d\lambda^n}\langle
e^{-\lambda\tau} \rangle \vert_{\lambda=0}$. Regardless of the choice of $a$
and $\rho(\tau)$, the characteristic functions $\mathcal{D}_\parallel$ and
$\mathcal{D}_\perp$ have a zero eigenvalue arising from the translation
invariance of Eq. \eqref{mean_field} \cite{note1}. There is a fold bifurcation as an eigenvalue of
$\mathcal{D}_\parallel$ crosses the origin when $a \mu_\tau = 1$, which marks
the disappearance of the translating state as seen  from
Eq. \eqref{speed_sqr}. Numerical analysis \cite{Engel} reveals an additional curve on the $(a, \
\mu_\tau)$ plane (below the curve $a\mu_\tau=1$) along which perturbations
parallel to the translation undergo a Hopf bifurcation as a complex pair of
eigenvalues of $\mathcal{D}_\parallel$ cross the imaginary axis.

As for perturbations perpendicular to the translational motion, there is another fold
bifurcation as an eigenvalue of $\mathcal{D}_{\perp}$ crosses the origin along
the curve $a\langle \tau^2\rangle = 2$, which represents a bifurcation in which the translating
state merges with a circularly rotating state of infinite radius, as
  discussed below.

Considering a fixed $\sigma_\tau$, the overall stability picture of the
translating state of Eqs. \eqref{mean_field} is as follows (see Fig. \ref{trans_bifs}). For values of $(a, \ \mu_\tau)$ below
the curves $a\langle\tau\rangle =1$ and $a\langle\tau^2\rangle =2$ (region
  A) the
translating state is linearly stable . These two curves may cross at a point
that we call the `zero frequency Hopf point' (ZFH). The transverse direction of the
translating state becomes unstable along the curve $a\langle\tau^2\rangle =2$
where this state merges with the circularly rotating state (along the
mentioned curve the rotating state has an infinite radius); transverse
  perturbations of the translating state will thus produce a transition
  to the rotating state in  regions B and C. From the ZFH
point, there emanates a Hopf bifurcation curve where the parallel component of
the translating state becomes unstable so that in region C there is a transition from the translating state to
  oscillations along a straight line. Finally, the translating state ceases
to exist along the curve $a\langle\tau\rangle =1$ where there is a pitchfork
type bifurcation with the stationary steady state solution. The possible
  behaviors in region D are discussed below.

\vspace{-0.3cm}

\begin{figure}[h]
\begin{minipage}{0.49\linewidth}
\includegraphics[width=4.3cm,height=3.0cm]{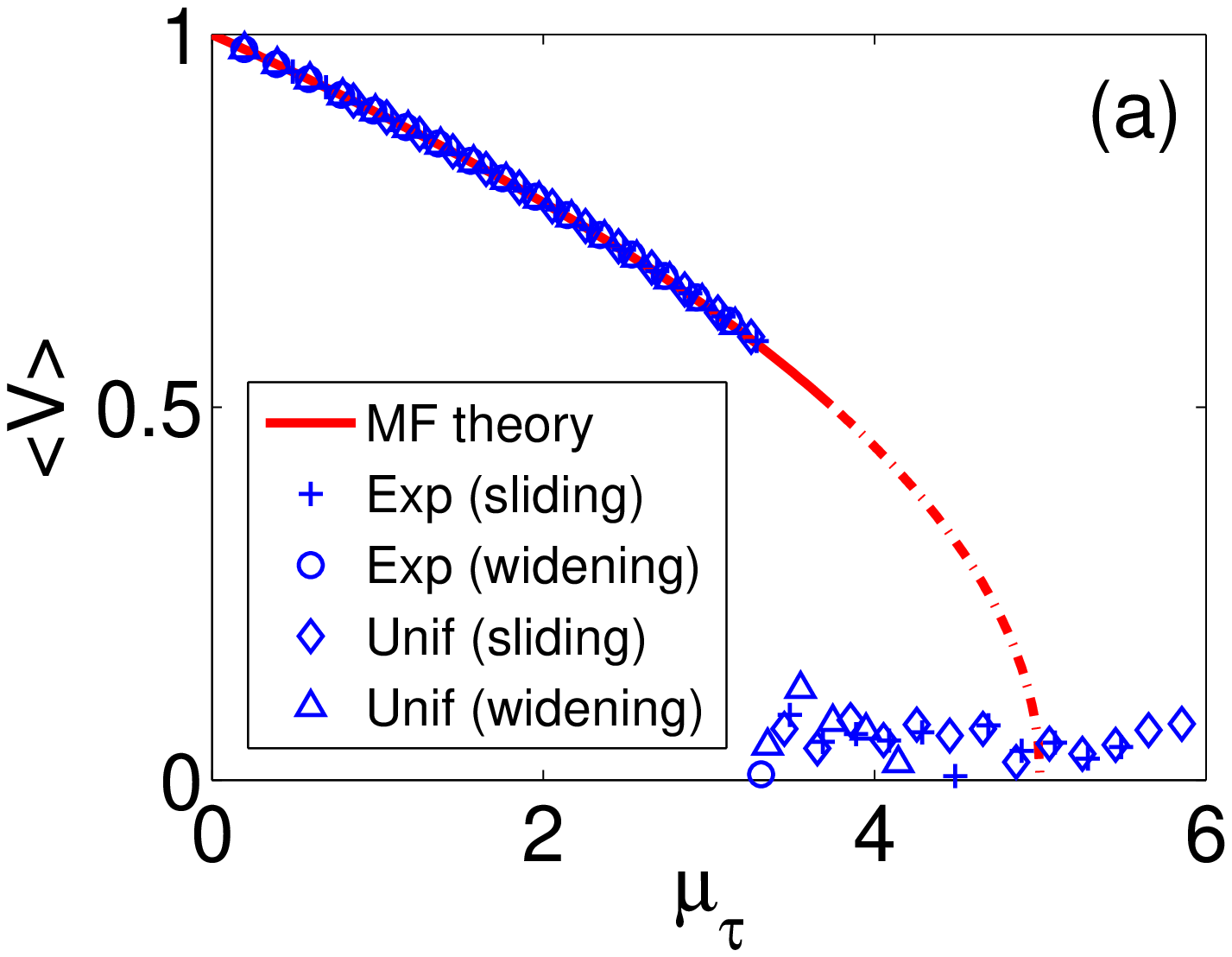}
\end{minipage}
\begin{minipage}{0.49\linewidth}
\includegraphics[width=4.3cm,height=3.0cm]{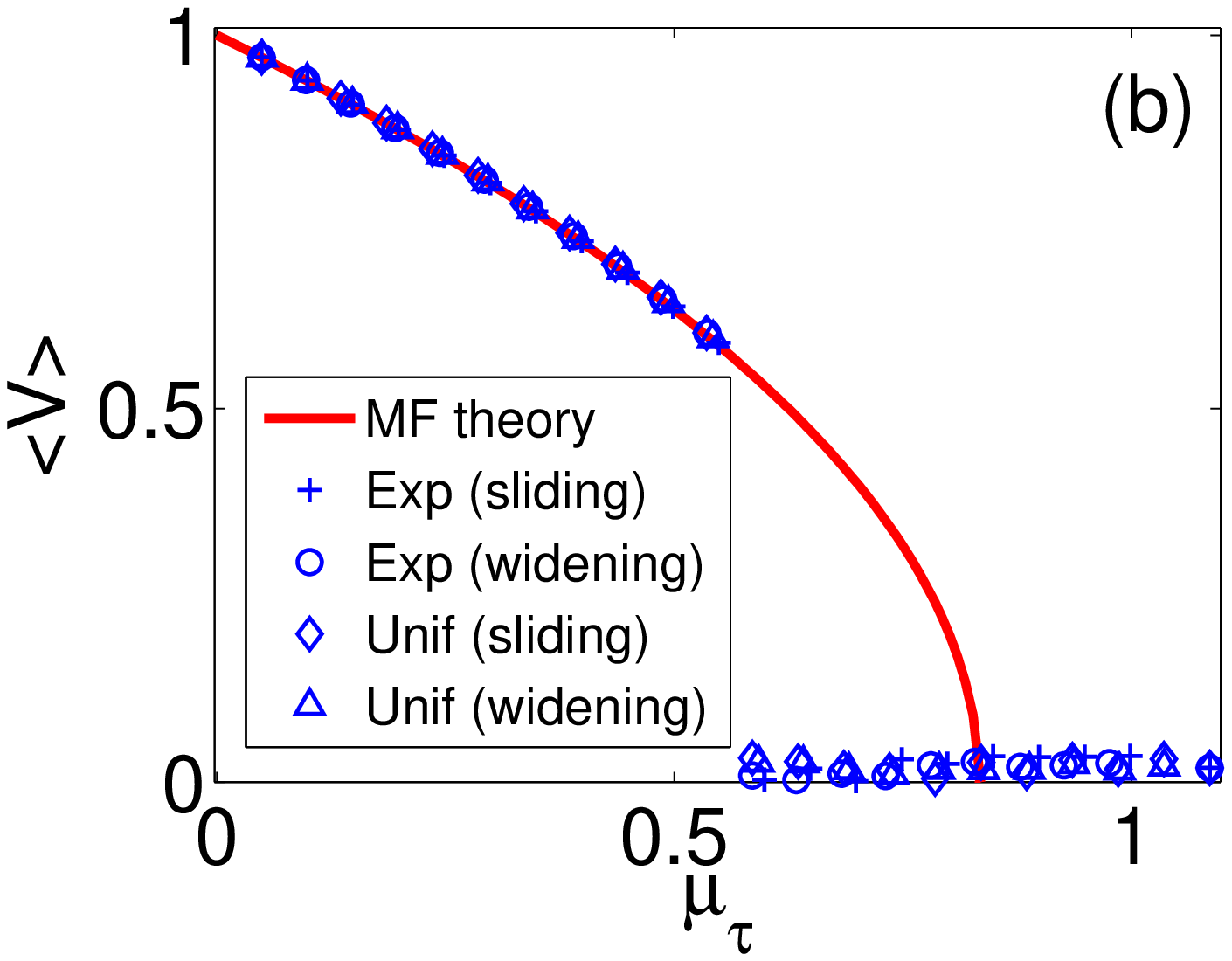}
\end{minipage}\\
\vspace{-0.3cm}
\caption{Speed of the translating state of Eqs. \eqref{swarm_eq} and
  \eqref{mean_field} as a function of $\mu_\tau$; here $N=150$, $a = 0.2$ (a)
  and $a=1.2$ (b). The red line (color online) represents the mean field result from
  Eq. \eqref{speed_sqr}; the continuous segment marks where the translating
  state is linearly stable and the dashed segment where it is unstable. The
  symbols represent numerical simulations of Eqs. \eqref{swarm_eq} for
  different time delay distributions: sliding exponential and sliding uniform $\sigma_\tau$ = 0.5
  (a),  $\sigma_\tau$ = 0.05 (b); widening exponential $\mu_\tau =
  \sigma_\tau$ (a) and (b); widening uniform $\mu_\tau =
  \sqrt{3} \sigma_\tau$ (a) and (b). }\label{trans_state_sim}
\end{figure}
\vspace{-0.2cm}

We compare the mean field bifurcation results with the full swarm system via numerical
simulations.  Here, we make use of two different time delay
distributions with mean $\mu_\tau$ and standard deviation $\sigma_\tau$ to test
our findings. The first is an exponential distribution $\rho(\tau) =
e^{\frac{\tau - \mu_\tau+\sigma_\tau}{\sigma_\tau}}/\sigma_\tau$ for $\tau \geq
\mu_\tau-\sigma_\tau$ and zero otherwise; we require $\sigma_\tau \leq
\mu_\tau$ for proper normalization. The second distribution is a uniform $\rho(\tau) = \frac{1}{2\sqrt{3} \cdot \sigma_\tau}$ for $\mu_\tau -
\sqrt{3}\sigma_\tau \leq \tau \leq \mu_\tau + \sqrt{3}\sigma_\tau$ and zero
otherwise; here, we require $\sqrt{3}\sigma_\tau \leq
\mu_\tau$. Moreover, we employ two versions of the mentioned distributions: a `sliding'
 one ($\sigma_\tau$ = const.) and a `widening' version ($\mu_\tau \propto \sigma_\tau$). 

 Fig. \ref{trans_state_sim} compares the speed of
the swarm obtained from Eq. \eqref{speed_sqr} with the time-averaged speed of
the center of mass obtained from simulations (after the decay of transients). The swarm particles are all
located at the origin at time zero and move with the speed obtained from
Eq. \eqref{speed_sqr} along the $x$ axis. In these simulations, we use both
`sliding' ($\sigma_\tau$ fixed) and `widening' (both $\mu_\tau$ and
$\sigma_\tau$ vary) versions of the exponential and uniform distributions
described above. Fig. \ref{trans_state_sim} shows that the swarm converges to
the translating state up to a value of $\mu_\tau$ beyond which the swarm
converges to a state in which it oscillates back and forth along a line with a
near zero time-averaged speed (the average is taken over an interval much
longer than the period). The transition to the oscillatory regime occurs
earlier than the mean field prediction. The full simulation results show
that in this deviation from the mean field, the swarm particles become spread
out too far apart and render the approximations leading to Eq. \eqref{mean_field} invalid.

\vspace{-0.3cm}

\begin{figure}[h]
\begin{minipage}{0.49\linewidth}
\includegraphics[width=4.3cm,height=3.0cm]{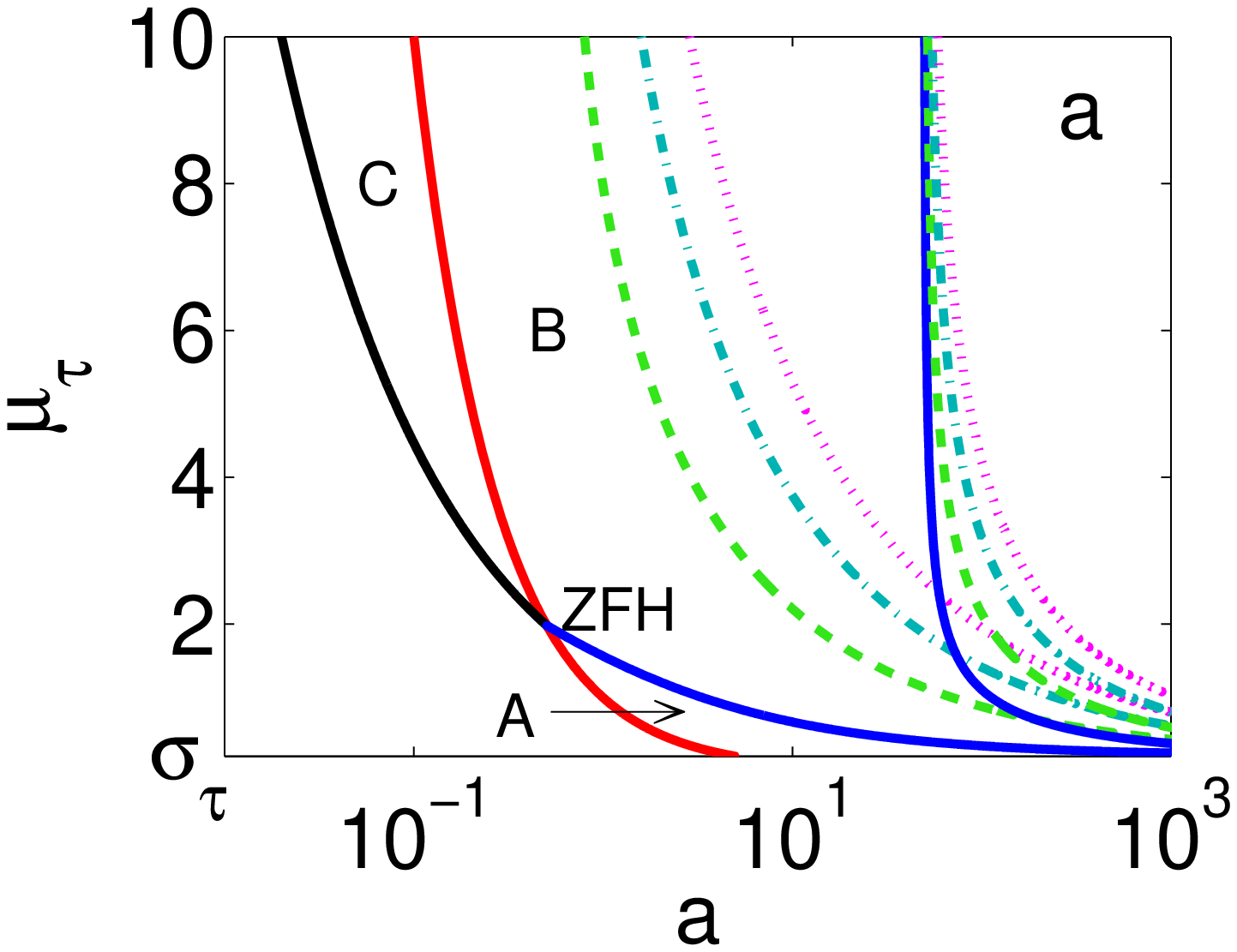}
\end{minipage}
\begin{minipage}{0.49\linewidth}
\includegraphics[width=4.3cm,height=3.0cm]{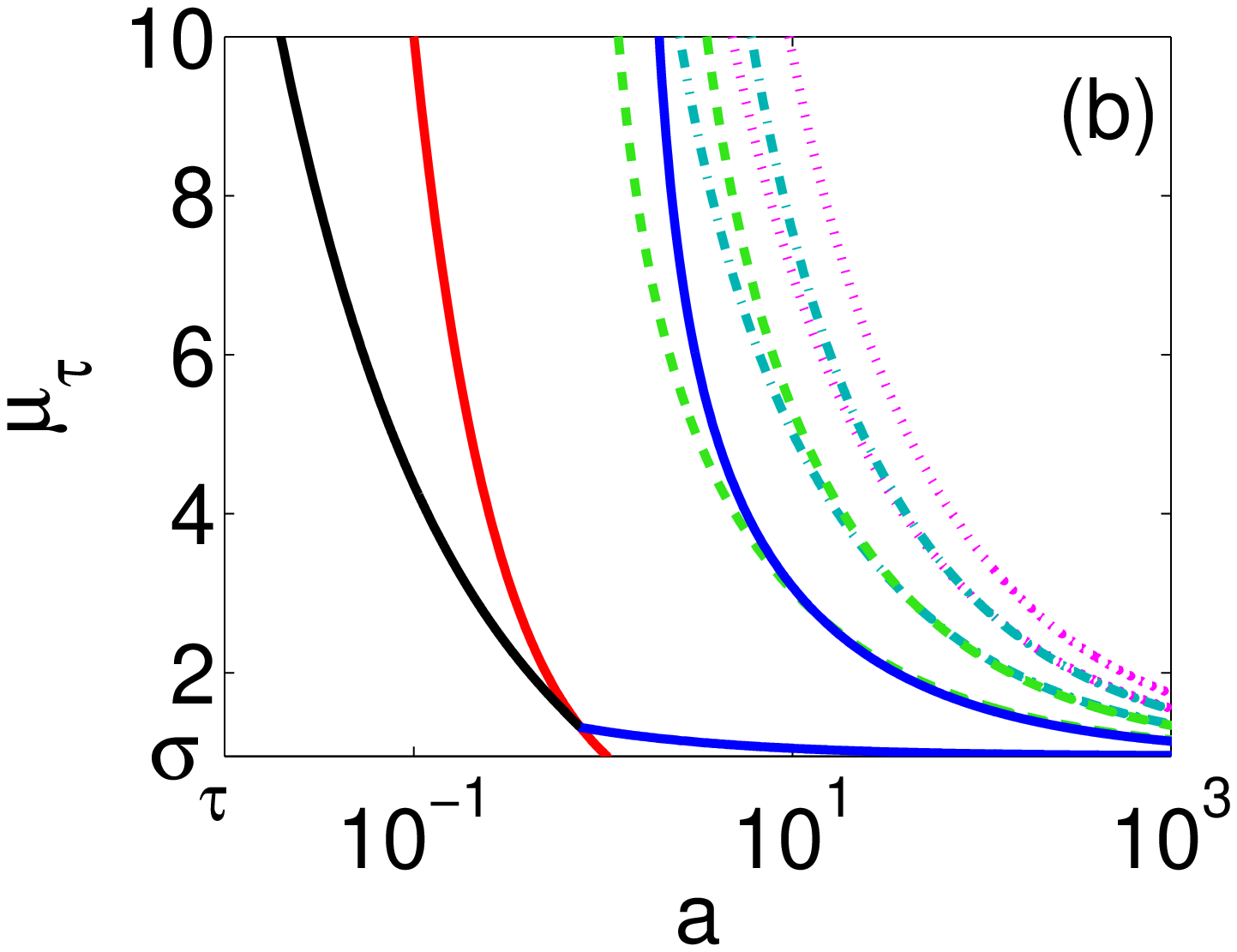}
\end{minipage}
\begin{minipage}{0.49\linewidth}
\includegraphics[width=4.3cm,height=3.0cm]{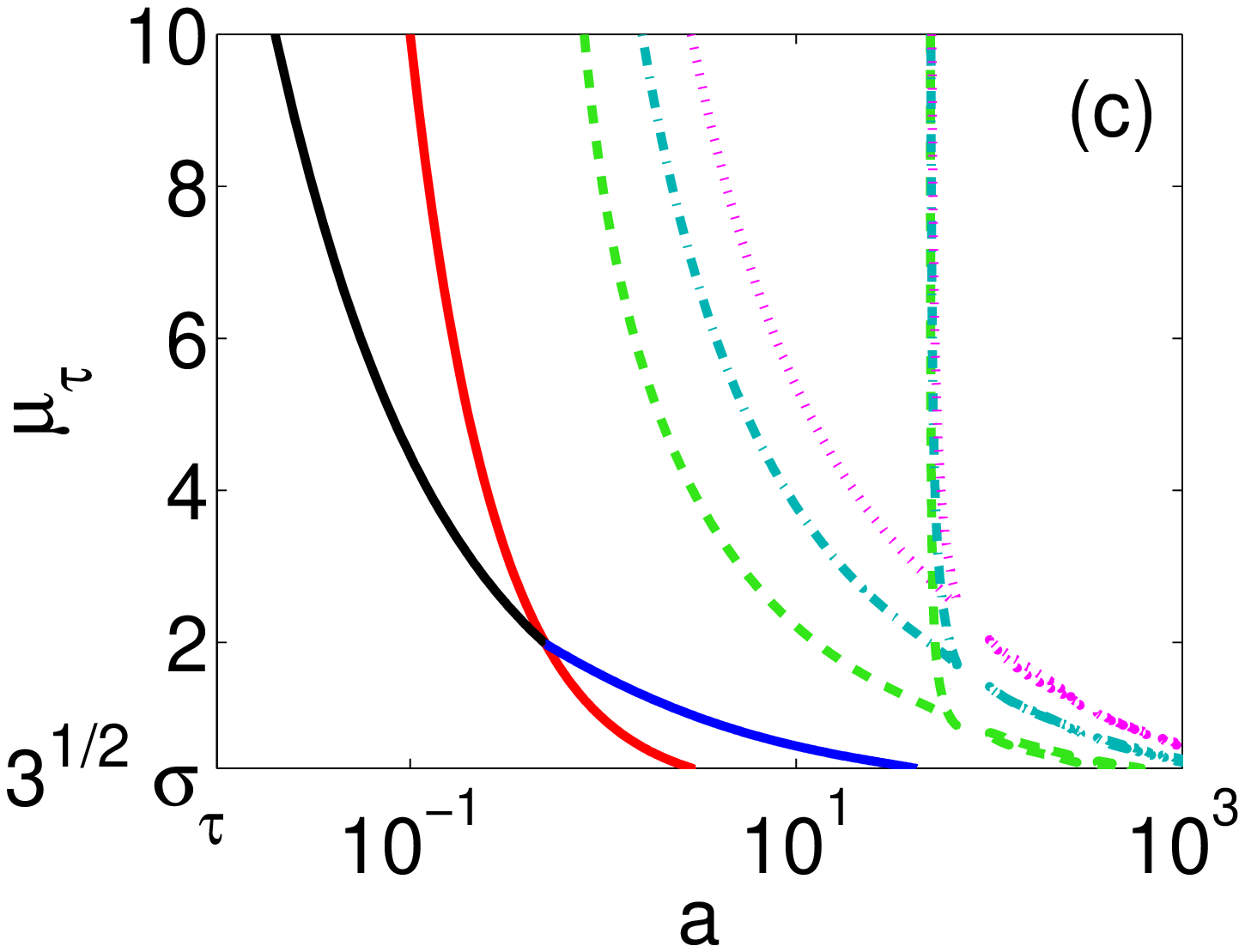}
\end{minipage}
\begin{minipage}{0.49\linewidth}
\includegraphics[width=4.3cm,height=3.0cm]{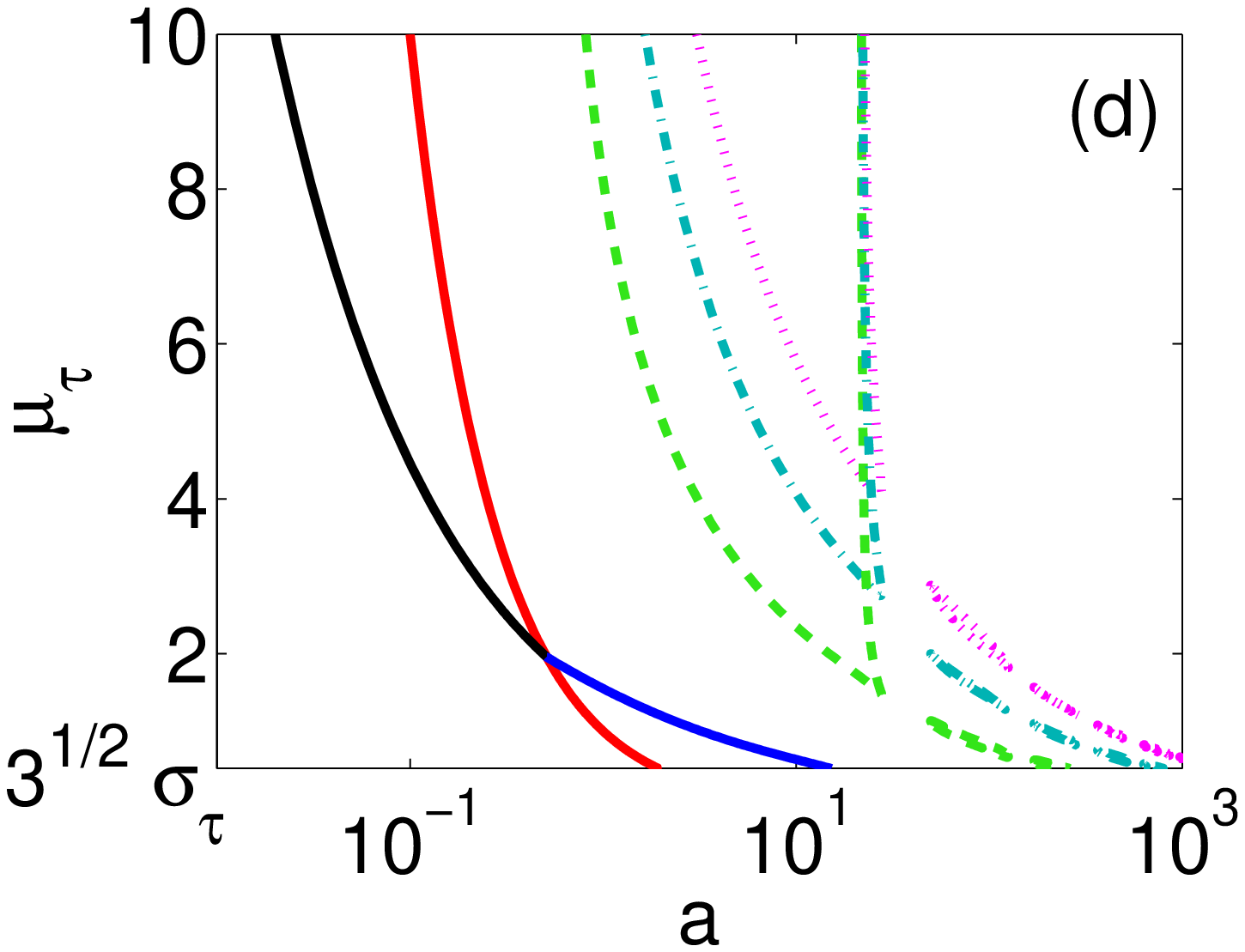}
\end{minipage}
\caption{Bifurcation curves of the mean field Eqs. \eqref{mean_field} at fixed
  $\sigma_\tau$ for the
  two  time delay distributions $\rho(
\tau)$  described in the text: exponential (top) and uniform (bottom). The translational state
disappearance curve $a\mu_\tau=1$ (red),  bifurcation of the translational
state with circularly rotating state curve $a \langle \tau^2 \rangle = 2$
(black). The first four members of the stationary state Hopf bifurcation
curves are also shown (blue, dashed green, dotted-dashed cyan and
dotted magenta). In each panel, $\sigma_\tau$ has the value (a) 0.2, (b) 0.95, (c) 0.2
and (d) 0.3,
respectively. (Color online.)}\label{stst_bifs}
\end{figure}

\vspace{-0.2cm}

In addition to the translating state, Eqs. \eqref{mean_field} always possess a stationary state
solution $\mathbf{R}(t) =\mathbf{R}_0=$ const. In the full system,
Eq. \eqref{swarm_eq}, the stationary state for the center of mass manifests itself in a
swarm `ring state', where some particles rotate clockwise and others
counter-clockwise on a circle around a static center of mass. The characteristic equation
that governs the linear stability of the stationary state has the form
$\big(\mathcal{D}(\lambda)\big)^2=0$, where $\mathcal{D}(\lambda) =
\mathcal{F}(\lambda) - \lambda.$ Once more there is a zero eigenvalue for all choices of $a$ and $\rho(\tau)$
that arises from the translation invariance of Eqs. \eqref{mean_field}. Also,
since $\mathcal{D}(0)= 0$, $\mathcal{D}'(0)=a\mu_\tau - 1$ and $\lim_{\lambda\rightarrow \infty}\mathcal{D}(\lambda) =
\infty$, the condition $a \mu_\tau - 1 < 0$ guarantees the existence of at
least one real and positive eigenvalue which renders the stationary state
linearly unstable. Thus, $a \mu_\tau = 1$ is a bifurcation curve on the $(a,\mu_\tau)$
plane along which the uniformly
translating state bifurcates with the stationary state.

The stationary state undergoes Hopf bifurcations when the equation
$\mathcal{D}(i \omega)= a\left(1 - \langle
  e^{-i\omega\tau}\rangle\right) -\omega^2 - i\omega = 0$ for $\omega \neq 0$ is satisfied. The function $\langle
e^{-i\omega\tau}\rangle$ is called the characteristic function of
$\rho(\tau)$ and is related to the moment generating function of the
distribution; its Taylor series contains all of the moments of
$\rho(\tau)$. This shows that the location of the Hopf bifurcations depends on
the values of all moments of
the time delay distribution. This is in  contrast to the region where the
translating state exists $a\mu_\tau < 1$, which involves the first moment only.

At the location of the Hopf bifurcations, circular orbits bifurcate from the
stationary state. This may be seen by changing Eq. \eqref{mean_field} from the
Cartesian $(X, \ Y)$ to polar coordinates $(R, \ \phi)$, noticing that
circular orbits $R = R_0$, $\phi = \omega t$ are possible as long as
\begin{align}\label{circ_orb}
\omega^2 = a\left(1 - \langle \cos\omega
  \tau\rangle\right), \qquad  R_0 = \frac{1}{\omega}\sqrt{1 - \frac{a}{\omega} \langle \sin\omega \tau\rangle },
\end{align}
and then realizing that the first of \eqref{circ_orb} and the condition
$R_0=0$ are precisely the real and imaginary parts of the Hopf bifurcation
conditions $\mathcal{D}(i\omega)=0$.

\vspace{-0.3cm}

\begin{figure}[h]
\begin{minipage}{0.49\linewidth}
\includegraphics[width=4.3cm,height=3.0cm]{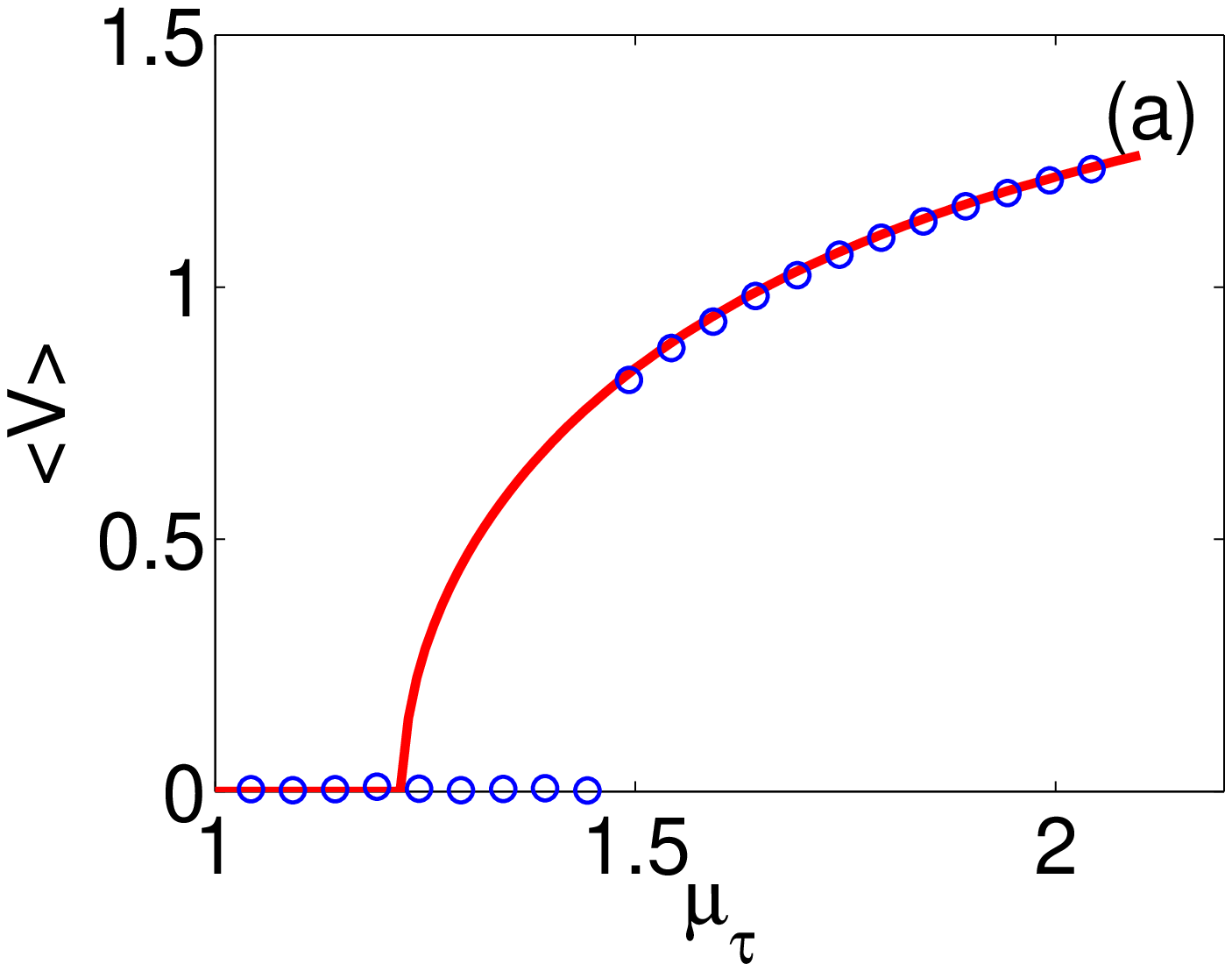}
\end{minipage}
\begin{minipage}{0.49\linewidth}
\includegraphics[width=4.3cm,height=3.0cm]{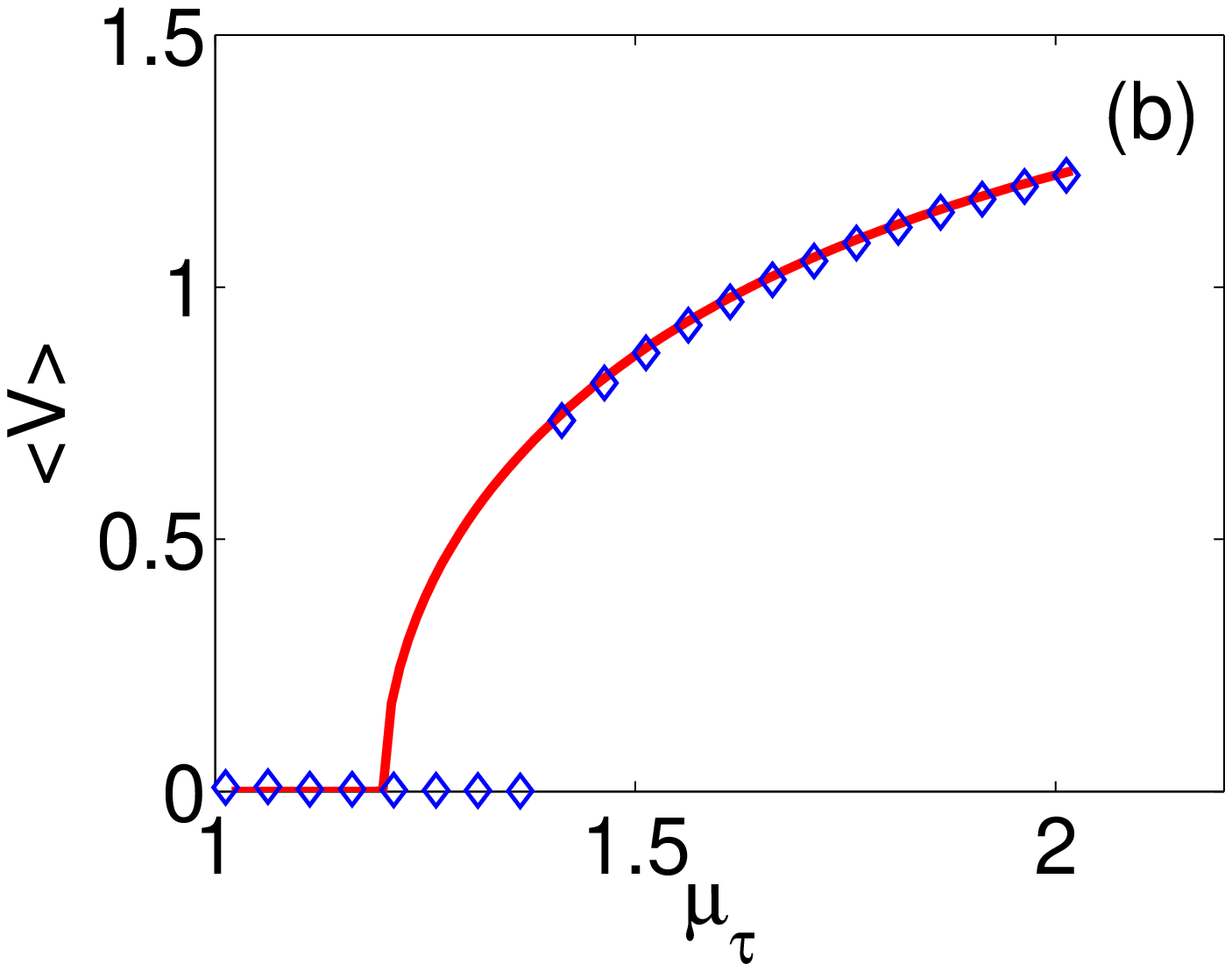}
\end{minipage}
\caption{Center of mass speed as a function of $\mu_\tau$ for the `sliding'
  exponential time delay distribution (left) and the `sliding' uniform
  distribution (right). Here $N=150$, $a = 2$. The continuous red curve  represents the mean field result from
  Eq. \eqref{circ_orb}, while the symbols represent the results from numerical simulations of
  Eqs. \eqref{swarm_eq}.  (color
  online)}\label{rot_state_sim}
\end{figure}

\vspace{-0.2cm}

Generically, the Hopf conditions for the stationary state
$\mathcal{D}(i\omega)=0$ yield a family of curves in the $(a, \ \mu_\tau)$
plane (Fig. \ref{stst_bifs}). The first member of the Hopf family emanates
from the crossing of the curves  $a\mu_\tau=1$ and $a \langle \tau^2\rangle =
2$; the former curve is where the translating state bifurcates from the
stationary state in a pitchfork-like bifurcation. Hence the name `Zero
Frequency Hopf' for the Hopf-fold point (Fig. \ref{stst_bifs}a). The first Hopf curve is supercritical
and gives rise to a circularly rotating orbit with radius and frequency given
by the first solution of Eqs. \eqref{circ_orb}. Below this first Hopf curve
  and $a\mu_\tau=1$, in region A the stationary state
  is stable. From Eqs. \eqref{circ_orb} it
follows that this circularly rotating orbit collides with the translating
state along the curve $a\langle \tau^2\rangle=2$ where its radius tends to
infinity and its speed to that of the translating state, $\sqrt{1 - a
  \mu_\tau}$. Thus, in regions B and C the system
  converges to the circularly rotating orbit. The different regions change
  shape for the other panels of Fig. \ref{stst_bifs}, but the dynamics remain
  as described above.

We  compare the mean field prediction for the location of the birth of the
circularly rotating state (first Hopf curve) with the full
system. Fig. \ref{rot_state_sim} shows the results from numerical simulations of
Eqs. \eqref{swarm_eq} at a fixed value of $\sigma_\tau$ for increasing
$\mu_\tau$. We plot the speed of the center of mass averaged over a long time
interval, after the decay of transients. In that plot, the near-zero values of the
mean speed (in the interval $0 \lesssim \mu_\tau \lesssim 1.4$) indicate that
the particles have converged to the ring state, while for all higher values of $\mu_\tau$ the swarm converges to the
rotating state. Remarkably, for the highest values of $\mu_\tau$,
the center of mass of the swarm moves faster than unit speed, the asymptotic speed of uncoupled particles.  The reason is that while in its rotating orbit, the
{mean time-}delayed position of the swarm is actually ``ahead'' of  the
  center of mass at the current time, causing the particles to
accelerate forward along the circular orbit.

In summary, we have considered a randomly delay distributed coupled swarm
model, and analyzed the bifurcations of various patterns as a function of
delay characteristics and coupling strength. In particular, we have shown that the location and shape of the Hopf
bifurcation curves is strongly-dependent on all the moments of
$\rho(\tau)$. This dependence, in addition to the fact that the succeeding Hopf curves in
Fig.~\ref{stst_bifs} exhibit higher frequencies of rotation, makes the
higher-order patterns equally sensitive to all the moments of the delay distribution.  In the single delay case with distribution  $\rho =
\delta(\tau - \tau_0)$, where $\delta(\tau)$ is a Dirac delta function, all of
the succeeding Hopf bifurcations are all subcritical and continuous. In contrast, when all
moments are present, the bifurcations may not even be continuous,  presenting
their structure as isolated closed curves bounded by fold bifurcations, as
seen in Fig.~\ref{stst_bifs}d for a uniform distribution. Finally, we expect that in other 
globally delay-coupled systems~\cite{Choi00, Marti03, Kozyreff01, Masoller09},  generic types of behavior involving
bifurcations including all moments of the distribution should be present.

The authors gratefully acknowledge the Office of Naval Research for their
support. LMR is an NIH post doctoral fellow and  BL is a post doctoral fellow of the NRC.  

\bibliographystyle{apsrev}

%\bibliography{/home1/nlschaos/luis/Documents/Bib_TeX/biblio,notes}

\begin{thebibliography}{31}
\bibitem[{not()}]{note1}
\bibinfo{note}{In fact, $\mathcal{D}_\perp$ has a second zero eigenvalue.}
\end{thebibliography}

\end{document}